\documentclass{ws-procs10x7}

\usepackage{graphicx}

\usepackage{balance}
\newcolumntype{d}[1]{D{.}{.}{#1}}

\begin{document}

\title{INDIRECT SEARCH FOR DARK MATTER WITH AMS IN POSITRONS,
       GAMMA AND ANTIPROTONS CHANNELS}

\author{D. CASADEI}

\address{INFN, Sezione di Bologna, Via Irnerio 46, 40126 Bologna, Italy\\
E-mail: Diego.Casadei@bo.infn.it}

\twocolumn[
\maketitle

\abstract{%
  The Alpha Magnetic Spectrometer (AMS), to be installed on the International
  Space Station, will provide data on cosmic radiations in a large energy
  range. The main physics goals in the astroparticle domain are the antimatter
  and the dark matter searches.  Dark matter should be composed of non
  baryonic weakly interacting massive particles, a good candidate being the
  lightest SUSY particle in R-parity conserving models.  As a prototype for
  the AMS-02 experiment, the AMS-01 particle spectrometer was flown on the
  Space Shuttle Discovery in near earth orbit for a ten day mission in June
  1998. The direct identification of positrons in AMS-01 was limited to
  energies below 3 GeV due to the vast proton background and the
  characteristics of the subdetectors, but the sensitivity towards higher
  energies (up to 40 GeV) was extended by identifying positrons through the
  conversion of bremsstrahlung photons.  AMS-02 will greatly improve the
  accuracy on the positron spectrum, which will be measured up to 300 GeV,
  together with the antiproton and $\gamma$-ray flux, thus providing a unique
  chance to measure all relevant neutralino decay channels with the same
  experiment.
}

]

\section{The AMS experiment}

The Alpha Magnetic Spectrometer (AMS) is a large-acceptance (0.4 m$^2$ sr)
space experiment to be operated for at least 3 years on board of the
International Space Station \cite{amsrep}, whose main goals are the search
for cosmic antimatter, for the signatures of the dark matter, for exotic
particles (like strangelets or heavy leptons), and the precise measurement of
cosmic ray (CR) rigidity
spectra from about 0.2 GV up to 2 TV.  The AMS international collaboration
developed a first version of the magnetic spectrometer (AMS-01) which was
flown on board of the space shuttle Discovery for a 10 days mission on June
1998 (NASA STS-91 flight), and is currently assembling an upgraded version of
the detector (AMS-02).

\begin{figure}[t]
\centering
\includegraphics[width=0.8\columnwidth]{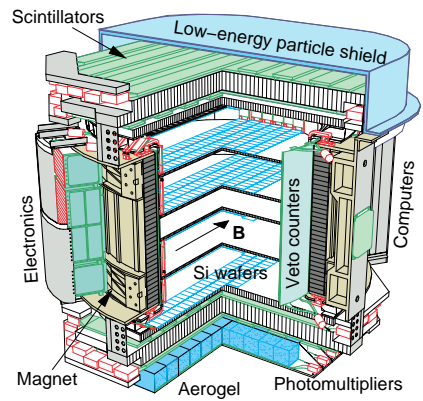}
\caption{The AMS-01 detector.}\label{ams01}
\end{figure}

The AMS-01 space spectrometer \cite{amsrep} (figure~\ref{ams01}) is based on a
permanent Ne-Fe-B magnet (dipolar field, 0.13 T maximum intensity) enclosing 6
planes of double-sided silicon tracker and the anticoincidence system (ACC),
consisting of 16 plastic scintillator paddles.  Above and below the magnet,
two couples of scintillator planes consisting of 14 counters each provide the
measurement of the time of flight (TOF) and an aerogel threshold \v{C}erenkov
counter (ATC) placed below the magnet is used to improve the separation
between positrons and protons.

\begin{figure}[t!]
\includegraphics[width=\columnwidth]{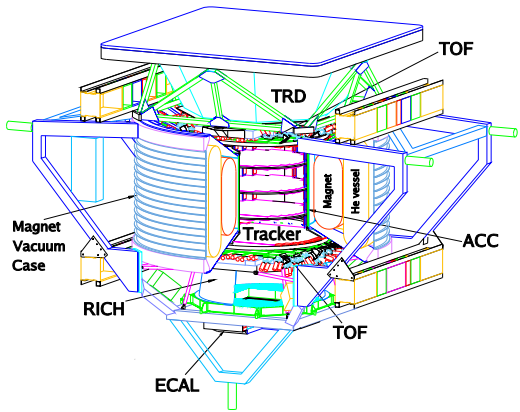}
\caption{The AMS-02 detector.}\label{ams02}
\end{figure}

The AMS-02 detector \cite{ams02} is based on a superconducting magnet (0.85 T
maximum intensity) enclosing the ACC system and 8 layers of double-sided Si
tracking planes with improved spatial resolution.  In addition, a
proximity-focusing ring-imaging \v{C}erenkov detector (RICH) replaces the ATC,
a transition-radiation detector (TRD) is positioned above the upper TOF planes
and an electromagnetic calorimeter (ECAL) is installed on the opposite side
(figure~\ref{ams02}).
With the help of the stronger magnetic field and the better tracking, the
maximum detectable rigidity
moves from about 0.2 TV to 2 TV, whereas the momentum resolution
goes from 5\% to 2\% in the range where the CR flux is maximum.

\begin{figure}[t!]
\includegraphics[width=\columnwidth]{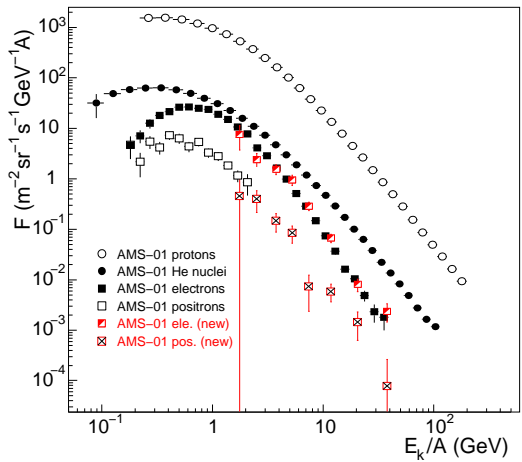}
\caption{Primary CR spectra measured by AMS-01.}\label{ams98}
\end{figure}

Figure \ref{ams98} shows the CR spectra of protons, He nuclei, electrons and
positrons measured by AMS-01 in June 1998
\cite{ams01pr,ams01el,ams01he,ams01elnew}.  The direct measurement of the
positron spectrum is limited by the high proton background (whose flux is
$\sim 10^4$ higher than positrons), which is suppressed with the help of the
ATC only below the \v{C}erenkov threshold (3 GeV).  At higher energy, a
different way has to be followed to suppress the background of a factor $\sim
10^{-6}$: a recent reanalysis of AMS-01 data \cite{ams01elnew} focused on the
rare events in which the primary electron or positron emitted a bremsstrahlung
photon which converted into a e$^+$-e$^-$ pair (resulting in 3 almost coplanar
tracks).  This has made possible to extend the energy range up to 40 GeV.

In addition to the measurement of the primary CR spectra, AMS-01 also studied
secondary particles, created by the interaction of CR in the Earth atmosphere.
A backtracing algorithm was used to follow all particles with energy below the
geomagnetic cut-off to see if they had originated in the atmosphere.  It was
found that the trapped particles may have short life before being absorbed
again in the atmosphere, or a relatively long life (longer than 1 s) when
bouncing back and forth between two mirror points.  AMS-01 found that this is
true for electrons and positrons \cite{ams01el}, protons \cite{ams01pr1}, and
helium \cite{ams01he}.

\section{Dark Matter}

When looking at the universe at large scales, several independent measurements
of the matter distribution (SDSS), of galactic and QSO red-shifts (2dF), of
the CMB spectrum (COBE, Boomerang, MAXIMA, WMAP) all agree on two important
cosmological features.  First, the universe is spatially flat, as a
consequence of the extraordinary expansion phase (``inflation'') which took
part in the very early stages of the cosmic evolution.  Second, the universe
dynamics are today vacuum dominated: we are living in a phase when the
accelerated cosmic expansion can be described as the effect of the
``cosmological constant'' $\Lambda$ of Einstein's equations, which is
responsible of about 74\% of the total energy density in the universe
\cite{tonry03}.

Though the energy density due to the matter content of the universe is quite
high (about 26\%), a number of observations (galaxy rotation curves, baryon to
photon ratio, light element abundances) put severe constraints on the amount
of baryons in the universe: they cannot constitute more than a few percent of
the total energy density \cite{wmap}.  The biggest fraction of the matter is
visible only thanks to its gravitational effects: it must be consisting of
neutral objects, hence the name of ``dark matter'' (DM).  In addition, DM
particles must have been not relativistic when leaving the radiation dominated
epoch, to be consistent with the power spectrum of the CMB and with the galaxy
formation, hence they are heavy.  Finally, beyond gravity, they may have only
weak interactions, hence the WIMP acronym (weakly interacting massive
particles).

Among the DM candidates (for example axions \cite{axion}, heavy leptons
\cite{heavylepton}, sterile neutrinos \cite{sterilenu}), the lightest
supersymmetric (SUSY) particles are the preferred option \cite{jungman96}.
WIMPs may be a mixture of the neutral SUSY weak-interaction eigenstates
(photino, wino and two higgsinos), the neutralinos $\chi$, whose mass $m_\chi$
is expected to be between several tens of GeV and few TeV.  The conservation
of R-parity implies that neutralinos and antineutralinos are stable.  In
addition, in the simplest SUSY models they are Majorana particles, i.e.\ they
are their own antiparticles.  Annihilations would produce fermion-antifermion
pairs (heavy fermions being strongly favored) or gauge bosons, whose end
products are stable particles: high-energy photons, electrons and positrons,
protons and antiprotons.  Hence, we might be able to detect the annihilation
signatures looking at the spectra of such particles.

The annihilation rate is $\Gamma = \langle \sigma v \rangle$, where $\langle
\sigma v \rangle$ is the thermally averaged total cross section for
annihilation times the relative velocity $v$, which is very small:
annihilations can be considered pratically at rest.  Neutralinos ceased to be
in thermal equilibrium with the other particles when $\Gamma < H$, the
expansion rate of the universe: at this point the annihilations were
exponentially suppressed and a relic cosmological abundance remained
\cite{jungman96}.  The current neutralino energy density, in units of the
crytical density $\rho_\mathrm{c} = 1.05 \times 10^{-5} h^2$ (where $h$ is the
Hubble parameter in units of 100 km s$^{-1}$ Mpc$^{-1}$), is \cite{jungman96}:
\begin{equation}
  \Omega_\chi h^2 = \frac{m_\chi n_\chi}{\rho_\mathrm{c}}
        \simeq \frac{3 \times 10^{-27} \mathrm{cm^3\, s^{-1}}}%
                    {\langle \sigma v \rangle}
\end{equation}
(independent from the mass $m_\chi$ apart from logarithmic corrections).  

A recent analysis of EGRET data \cite{deboer06} has shown that the diffuse
hig-energy gamma-ray spectrum is consistent with a model of galactic CR
propagation in which the annihilations of neutralinos with mass $m_\chi$ =
50--100 GeV is considered.  This result has been debated \cite{bergstrom06},
because the same model seems to overproduce antiprotons and requires a
peculiar DM distribution for our Galaxy.  To obtain convincing results, we
must observe the effects of WIMPs annihilations in all decay channels.  The
best ones are the antiprotons and positrons at high energy, because proton and
electron spectra are dominated by the ordinary cosmic rays.  In particular,
one would expect some excess in the positron/electron and antiproton/proton
ratios around the mass of the neutralinos.

\section{AMS sensitivity to DM signatures}

\begin{figure}[t!]
\centering
\includegraphics[width=0.8\columnwidth]{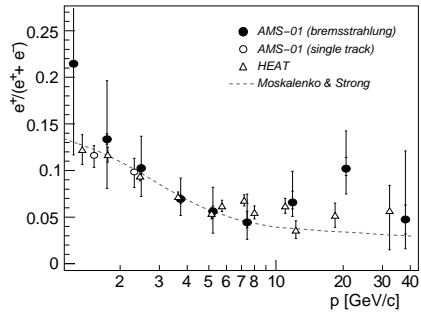}
\caption{Positron fraction measured by AMS-01 and HEAT.}\label{e+e-ratio}
\end{figure}

\begin{figure}[t!]
\centering
\includegraphics[width=0.8\columnwidth]{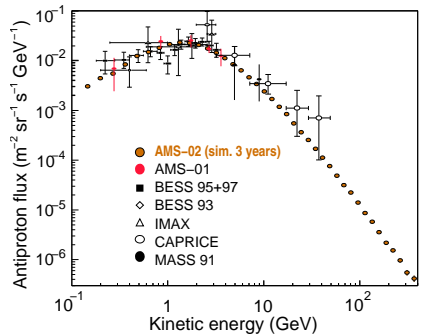}
\caption{Antiproton flux.  Expected results from AMS-02 are shown together
  with existing measurements.}\label{antip}
\end{figure}

Though the positron fraction measured by AMS and HEAT (figure
~\ref{e+e-ratio}, from ref.\cite{ams01elnew}) shows a possible excess at high
energy with respect to the ordinary CR component (dashed line
\cite{moskalenko98}), still no conclusive evindence has been found about the
possibility that neutralino annihilations really contribute to the
observed spectra.  In particular, the high energy part of the antiproton
spectrum is still uncovered (figure~\ref{antip}).

\begin{figure}[t!]
\centering
\includegraphics[width=0.8\columnwidth]{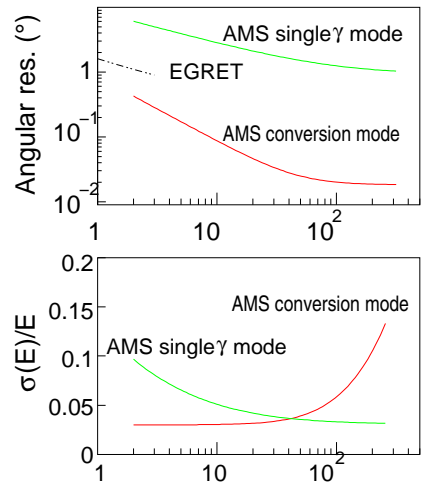}
\caption{Gamma-ray sensitivity of AMS-02.}\label{ams-gammas}
\end{figure}

The AMS-02 detector will be able to accurately measure the spectrum of
protons, antiprotons, electrons, positrons and $\gamma$-rays up to 300 GeV: a
unique possibility to measure all relevant decay channels with the same
experiment.  Figure~\ref{ams-gammas} shows the AMS-02 angular resolution (top
panel) and energy resolution (bottom panel) for $\gamma$-rays
\cite{jacholkowska06}.  Figure~\ref{antip} shows the expectation after 3 years
for the the antiproton spectrum, for the null hypothesis (no WIMP
annihilation).  The statistical errors are comparable to the symbol size:
AMS-02 will be able to check for DM annihilation signatures with unprecedented
sensitivity.

\end{document}